\documentclass[12pt]{article}
\usepackage{epsfig,psfrag,verbatim,amssymb,amsfonts,amsthm}
\usepackage{graphicx}
\usepackage{subfigure,url,amsmath}
\usepackage{subfigure,amsmath}
\usepackage{multirow}
\usepackage{color}
\usepackage{natbib}
\usepackage{vmargin} 
\usepackage{url}
\usepackage{authblk}

\usepackage{graphicx}
 \usepackage{mathptmx}      
\usepackage{epsfig,psfrag,verbatim,amssymb,amsfonts}
\usepackage{url,amsmath}
\usepackage{multirow}
\usepackage{color}
\usepackage{natbib}
\usepackage{enumerate}

\setpapersize{USletter}
\setmarginsrb{1truein}{1.1truein}{1truein}{1.3truein}{0pt}{0pt}{0pt}{0pt}
\def\bs{{\bf s}}

\def\bA{{\bf A}}
\def\bB{{\bf B}}

\def\bD{{\bf D}}

\def\bH{{\bf H}}
\def\bI{{\bf I}}

\def\bK{{\bf K}}

\def\bP{{\bf P}}

\def\bS{{\bf S}}

\def\bU{{\bf U}}
\def\bV{{\bf V}}

\def\bX{{\bf X}}
\def\bY{{\bf Y}}

\def\thick#1{\hbox{\rlap{$#1$}\kern0.25pt\rlap{$#1$}\kern0.25pt$#1$}}

\def\bepsilon{{\thick\epsilon}}

\def\bxi{{\thick\xi}}

\def\bpsi{{\thick\psi}}

\def\bSigma{{\thick\Sigma}}

\def\bPsi{{\thick\Psi}}

\def\smbalpha{{\thick{\scriptstyle{\alpha}}}}

\def\bAhat{{\widehat \bA}}

\def\bKhat{{\widehat \bK}}

\def\bYhat{{\widehat \bY}}

\def\bAtilde{{\widetilde \bA}}

\def\bKtilde{{\widetilde \bK}}

\def\bYtilde{{\widetilde \bY}}

\def\lambdahat{{\widehat\lambda}}

\def\xihat{{\widehat\xi}}

\def\sigmahat{{\widehat\sigma}}

\def\psihat{{\widehat\psi}}

\def\bxihat{{\widehat\bxi}}

\def\bSigmahat{{\widehat\bSigma}}

\def\smbalpha{\widehat{\smbalpha}}

\def\hbar{{\overline h}}

\def\inv{^{-1}}

\def\digamma{\mbox{digamma}}

\def\tr{\mbox{tr}}

\def\beq{\begin{equation}}
\def\eeq{\end{equation}}
\def\bev{\begin{verbatim}}
\def\eev{\end{verbatim}}

\def\lboxit#1{\vbox{\hrule\hbox{\vrule\kern6pt
      \vbox{\kern6pt#1\kern6pt}\kern6pt\vrule}\hrule}}

\def\thickboxit#1{\vbox{{\hrule height 1mm}\hbox{{\vrule width 1mm}\kern6pt
          \vbox{\kern6pt#1\kern6pt}\kern6pt{\vrule width 1mm}}
               {\hrule height 1mm}}}

\def\beq{\begin{eqnarray}}
\def\eeq{\end{eqnarray}}
\def\beqn{\begin{eqnarray*}}
\def\eeqn{\end{eqnarray*}}
\def\bse{\begin{eqnarray*}}
\def\ese{\end{eqnarray*}}
\def\raybe{\begin{eqnarray}}
\def\rayee{\end{eqnarray}}

\def\fat#1{\hbox{\rlap{$#1$}\kern0.25pt\rlap{$#1$}\kern0.25pt$#1$}}

\newtheorem{thm}{Theorem}
\newtheorem{prop}{Proposition}

\def\tr{{\textrm tr}}

\begin{document}
\newcommand*\samethanks[1][\value{footnote}]{\footnotemark[#1]}
\author{\normalsize Luo Xiao\thanks{Department of Biostatistics, Bloomberg School of Public Health,
Johns Hopkins University, Baltimore, MD},
David Ruppert \thanks{Department of Statistical Science and School of Operations Research and Information Engineering, Cornell University, Ithaca, NY},
Vadim Zipunnikov\samethanks[1],
and
Ciprian Crainiceanu\samethanks[1]
}
\title{{\Large Fast Covariance Estimation for High-dimensional Functional Data}}
\date{\today}
\maketitle

\begin{abstract} We propose two fast covariance smoothing methods and
associated software that scale up linearly with the number of
observations per function.
Most available methods and software cannot smooth covariance
matrices of dimension $J>500$; a recently introduced sandwich
smoother is an exception but is not adapted to smooth covariance
matrices of  large dimensions, such as $J= 10,000$. We introduce two
new methods that circumvent those problems: 1) a  fast
implementation of the sandwich smoother for covariance smoothing;
and 2) a two-step procedure that first obtains the singular value
decomposition of the data matrix and then smoothes the eigenvectors.
These new approaches are at least an order of magnitude faster in
high dimensions and drastically reduce computer memory requirements.
The new approaches provide instantaneous (a few seconds) smoothing
for matrices of dimension $J=10,000$ and very fast ($<$ 10 minutes)
smoothing for $J=100,000$.
 R functions, simulations, and data analysis provide ready to use,
reproducible, and scalable tools for practical data analysis of
noisy high-dimensional functional data.\\

\end{abstract}

\section{Introduction}
The covariance function plays an important role in functional
principal component analysis (fPCA), functional linear regression,
and functional canonical correlation analysis (see, e.g.,
\citealt{Ramsay:02, Ramsay:05}). The major difference between the
covariance function of functional data  and the covariance matrix of
multivariate data is that functional data is measured on the same
scale, with sizable noise and possibly sampled at an irregular grid.
Ordering of functional observations is  also important, but it can
easily be handled by careful indexing.  Thus, it has become common
practice in functional data analysis to estimate functional
principal components by diagonalizing a smoothed estimator of the
covariance function; see, e.g.,
\cite{Besse:86,Ramsay:91,Kneip:94,Besse:97,Staniswalis:98,Yao:03,Yao:05}.

Given a sample of functions,  a simple estimate of the covariance
function is the sample covariance. The sample covariance, its
eigenvalues and eigenvectors have been shown to converge to their
population counterparts at the optimal rate when the sample paths
are completely observed without measurement error
(\citealt{Dauxois:82}). However, in practice, data are measured at a
finite number of locations and often with sizable measurement error.
For such data the eigenvectors of the sample covariance matrix tend
to be noisy, which can substantially reduce interpretability.
Therefore, smoothing is often used to estimate the functional
principal components; see, e.g.,
\cite{Besse:86,Ramsay:91,Rice:91,Kneip:94,Capra:97,Besse:97,Staniswalis:98,Cardot:00,Yao:03,Yao:05}.
There are three main approaches to estimating smooth functional
principal components. The first approach is to smooth the functional
principal components of the sample covariance function; for a
detailed discussion see, for example,
\cite{Rice:91,Capra:97,Ramsay:05}. The second  is to smooth the
covariance function and then diagonalize it; see, e.g.,
\cite{Besse:86,Staniswalis:98,Yao:03}. The third  is to smooth each
curve and diagonalize the sample covariance function of the smoothed
curves; see \cite{Ramsay:05} and the references therein. Our first
approach is a fast bivariate smoothing method for the covariance
operator which connects the latter two approaches. This method is a
fast and new implementation of the  `sandwich smoother' in
\cite{Xiao:12}, with a completely different and specialized
computational approach that improves the original algorithm's
computational efficiency  by at least an order of magnitude. The
sandwich smoother with the new implementation
 will be referred to as Fast Covariance Estimation, or
FACE. Our second approach is to use smoothing spline smoothing of
the eigenvectors obtained from a high-dimensional singular value
decomposition of the raw data matrix and will be referred to as
smooth SVD, or SSVD. To the best of our knowledge, this approach has
not been used in the literature for low- or high-dimensional data.
Given the simplicity of SSVD, we will focus more on FACE, though
simulations and data analysis will be based on both approaches.

The  sandwich smoother  provides  the next level of computational
scalability for bivariate smoothers and has significant
computational advantages over  bivariate \textsl{P}-splines
(\citealt{Eilers:03,Marx:05}) and thin plate regression splines
(\citealt{Wood:03}). This is achieved, essentially, by transforming
the technical problem of bivariate smoothing into a short sequence
of univariate smoothing steps. For covariance matrix smoothing, the
sandwich smoother was shown to be much faster than local linear
smoothers. However, adapting the sandwich smoother to fast
covariance matrix smoothing in the ultrahigh dimensions of, for
example, modern medical imaging or high density wearable sensor
data, is not straightforward. For instance, the sandwich smoother
requires the sample covariance matrix which can be  hard to
calculate and impractical to store for ultrahigh dimensions. While
the sandwich smoother is the only available fast covariance
smoother, it was never tested for dimensions $J>5,000$ and becomes
computationally impractical for $J>5,000$ on current standard
computers. All of these dimensions are well within the range of
current high-dimensional data.

In contrast, our novel approach, FACE, is linear in the number of
functional observations per subject, provides instantaneous ($<$ 1
minutes)
 smoothing for matrices of dimension $J=10,000$ and
fast ($<$ 10 minutes) smoothing for $J=100,000$. This is done by
carefully exploiting the low-rank structure of the sample
covariance, which allows smoothing and spectral decomposition of the
smooth estimator of the covariance {\it without calculating or
storing the empirical covariance operator}. The new approach is at
least an order of magnitude faster in high dimensions and
drastically reduces memory requirements; see Table~\ref{time} in
Section~\ref{sec::sim} for a comparison of computation time. Unlike
the sandwich smoother, FACE also efficiently estimates the
covariance function, eigenfunctions, and scores.

The remainder of the paper is organized as follows.
Section~\ref{sec::settings} provides the model and data structure.
Section~\ref{sec::face} introduces FACE and provides the associated
fast
 algorithm.  Section~\ref{sec::extension}
 extends FACE to structured high-dimensional functional data and  incomplete data. Section~\ref{sec::alternative} introduces SSVD, the smoothing spline smoothing of eigenvectors obtained from SVD.  Section~\ref{sec::sim} provides simulation
results. Section~\ref{sec::example} shows how FACE works in a large
study of sleep. Section~\ref{sec::dis} provides concluding remarks.

FACE and SSVD are now implemented as  R functions ``fpca.face" and
``fpca2s", respectively, in the publicly available package {\it
refund} (\citealt{Crainiceanu:13}).


\section{Model and data structure}\label{sec::settings}
Suppose that $\{X_i, i=1,\dots, I\}$  is a collection of independent
realizations of a random functional process $X$ with covariance
function $K(s,t), s, t\in [0,1]$. The observed data, $ Y_{ij} =
X_i(t_j) + \epsilon_{ij}$, are noisy proxies of $X_i$ at the
sampling points $\{t_1, . . . , t_J\}$. We assume that
$\epsilon_{ij}$ are i.i.d.\ errors with mean zero and variance
$\sigma^2$, and are mutually independent of the processes $X_i$.

The sample covariance function can be computed  at each pair of
sampling points $(t_j, t_{\ell})$ by $\widehat{K}(t_j, t_{\ell}) =
I^{-1}\sum_i Y_{ij}Y_{i\ell}$. For ease of presentation we assume
that $Y_{ij}$ have been centered across subjects. The sample
covariance matrix, $\bKhat$,  is the $J\times J$ dimensional matrix
with the $(j,\ell)$ entry equal to $\widehat{K}(t_j, t_{\ell})$.
Covariance smoothing typically refers to applying bivariate
smoothers to $\bKhat$. Let $\mathbf{Y}_i = (Y_{i1},\dots, Y_{iJ})^T,
i =1,\dots, I$, then $\bKhat = I^{-1}\sum_{i=1}^I
\mathbf{Y}_i\mathbf{Y}_i^T = I^{-1}\mathbf{Y}\mathbf{Y}^T$,  where
$\mathbf{Y} = [\mathbf{Y}_1,\dots, \mathbf{Y}_I]$ is a $J\times I$
dimensional matrix with the $i$th column equal to $\mathbf{Y}_i$.
When $I$ is much smaller than $J$, $\bKhat$ is of low rank; this
low-rank structure of $\bKhat$ will be particularly useful for
deriving fast methods for smoothing $\bKhat$.

\section{FACE}\label{sec::face}
The FACE estimator of the covariance matrix has the following form
\beq \label{Ktilde} \bKtilde = \bS\bKhat\bS, \eeq where $\bS$ is a
symmetric smoother matrix of dimension $J\times J$. Because
of~(\ref{Ktilde}), we say FACE has a sandwich form. We use
\textsl{P}-splines (\citealt{Eilers:96}) to construct $\bS$ so that
$ \mathbf{S}= \mathbf{B}\left(\mathbf{B}^T\mathbf{B}+\lambda
\mathbf{P}\right)^{-1}\mathbf{B}^T$. Here $\mathbf{B}$ is the
$J\times c$ design matrix $\{B_k(t_j)\}_{1\leq j\leq J, 1\leq k\leq
c}$,   $\mathbf{P}$ is a symmetric penalty matrix of size $c\times
c$, $\lambda$ is the smoothing parameter,
$\{B_1(\cdot),\dots,B_c(\cdot)\}$ is the collection of B-spline
basis functions,  $c$ is the number of interior knots plus the order
(degree plus 1) of B-splines. We assume that the knots are equally
spaced and use a difference penalty as in \cite{Eilers:96} for the
construction of $\mathbf{P}$.  Model~(\ref{Ktilde})  is a special
case of the sandwich smoother in \cite{Xiao:12} as the two smoother
matrices for FACE are identical. However, FACE is specialized to
smooth covariance matrices and has some further important
characteristics.

First, $\bKtilde$ is guaranteed to be symmetric and positive
semi-definite because $\bKhat$ is so. Second, the sandwich form of
the smoother and the low-rank structure of the sample covariance
matrix can be exploited to scale FACE to high and ultra high
dimensional data ($J>10,000$).
For instance, the eigendecomposition of $\bKtilde$ provides the
estimates of the eigenfunctions associated with the covariance
function. However, when $J$ is large,  both the smoother matrix and
the sample covariance matrix are  high dimensional and even storing
them may become impractical.  FACE, unlike the sandwich smoother, is
designed to obtain the eigendecomposition of $\bKtilde$ {\it without
computing the smoother matrix or the sample covariance matrix}.

FACE depends on a single smoothing parameter, $\lambda$, which needs
to be selected. The algorithm for selecting $\lambda$ in
\cite{Xiao:12} requires $O(J^2I)$ computations and can be hard to
compute when $J$ is large. We propose efficient smoothing parameter
estimation algorithms that requires only $O(JIc)$ computations; see
Section~\ref{sec::gcv} for details.
\subsection{Estimation of eigenfunctions}\label{sec::spectral}
Assuming that the covariance function $K$ is in $L_2([0,1]^2)$,
Mercer's  theorem states that $K$ admits an eigendecomposition
$K(s,t) = \sum_k \lambda_k\psi_k(s)\psi_k(t)$ where
$\{\psi_k(\cdot): k\geq 1\}$ is a set of orthonormal basis of
$L_2([0,1])$ and $\lambda_1\geq\lambda_2\geq\cdots$ are the
eigenvalues. Estimating the functional principal
components/eigenfunctions $\psi_k$'s is one of the most fundamental
tasks in functional data analysis and has attracted a lot of
attention  (see, e.g., \citealt{Ramsay:05}). Typically,  interest
lies in seeking the first few eigenfunctions that explain a large
proportion of the observed variation. This is equivalent to finding
the first few eigenfunctions whose linear combination could well
approximate the random functions $X_i$. Computing the eigenfunctions
of a symmetric bivariate function is generally not trivial. The
common practice is to  discretize the estimated covariance function
and approximate its eigenfunctions by the corresponding eigenvectors
(see, e.g., \citealt{Yao:03}). In this section, we show that by
using FACE we can easily obtain the eigendecomposition of the
smoothed covariance matrix $\bKtilde$ in equation~(\ref{Ktilde}).

We start with  the decomposition
$(\mathbf{B}^T\mathbf{B})^{-1/2}\mathbf{P}(\mathbf{B}^T\mathbf{B})^{-1/2}
 = \mathbf{U}\textrm{diag}(\mathbf{s})\mathbf{U}^T$, where
$\mathbf{U}$ is the matrix of eigenvectors and $\mathbf{s}$ is the
vector of eigenvalues. Let $\mathbf{A}_S =
\mathbf{B}(\mathbf{B}^T\mathbf{B})^{-1/2}\mathbf{U}. $ Then
$\mathbf{A}_S^T\mathbf{A}_S = \mathbf{I}_{c}$ which implies that
$\mathbf{A}_S$ has orthonormal columns.  It follows that $
\mathbf{S} = \mathbf{A}_S\bSigma_S\mathbf{A}_S^T $ with $ \bSigma_S
= \left\{\mathbf{I}_{c}+\lambda
\textrm{diag}(\mathbf{s})\right\}^{-1}$. Let $\bYtilde= \bA_S^T\bY$
be a $c\times I$ matrix, then $ \bKtilde=
\bA_S\left(I^{-1}\bSigma_S\bYtilde
\bYtilde^T\bSigma_S\right)\bA_S^T. $ Thus only the $c\times c$
dimensional matrix in the parenthesis depends on the smoothing
parameter; this observation will lead to a simple spectral
decomposition of $\bKtilde$. Indeed, consider the spectral
decomposition $ I^{-1}\bSigma_S\bYtilde\bYtilde^T\bSigma_S =
\bA\bSigma\bA^T$, where $\bA$ is the $c\times c$ matrix of
eigenvectors and $\bSigma$ is the $c\times c$ diagonal matrix of
eigenvalues.  It follows  that $ \bKtilde =
(\bA_S\bA)\bSigma(\bA_S\bA)^T $ which is the eigendecomposition of
$\bKtilde$ and shows that $\bKtilde$ has no more than $c$ nonzero
eigenvalues (Proposition~\ref{prop1}). Because of the dimension
reduction of matrices ($c\times c$ versus $J\times J$), this
eigenanalysis of the  smoothed covariance matrix is fast.  The
derivation reveals that through smoothing we obtain a smoothed
covariance operator and its associated eigenfunctions. An important
consequence is that the number of elements stored in  memory is only
$O(Jc)$ for FACE, while using other bivariate smoothers requires
storing the $J\times J$ dimensional covariance operators. This makes
a dramatic difference, allows non-compromise smoothing of covariance
matrices, and provides a transparent, easy to use method.

\subsection{Selection of the smoothing parameter}\label{sec::gcv}
We start with the following result.
\begin{prop}
\label{prop1} Assume $c = o(J)$, then the rank of the smoothed
covariance matrix $\bKtilde$ is at most $\min(c,I)$.
\end{prop}
This indicates that  the number of knots controls the maximal rank
of the smoothed covariance matrix, $\bKtilde$,  or equivalently, the
number of eigenfunctions that can be extracted from $\bKtilde$. This
implies that using an insufficient number of knots may result in
severely biased estimates of eigenfunctions and number of
eigenfunctions. We propose to use a relatively large number of
knots, e.g., 100 knots, to reduce the estimation bias and control
overfitting by an appropriate penalty. Note that for
high-dimensional data, $J$ can be thousands or more and the
dimension reduction by FACE is sizeable.  Moreover, as only a small
number of functional principal components is typically  used in
practice, FACE with 100 knots seems adequate for most applications.
When the covariance function has a more complex structure or a
larger number of functional principal components are needed, one may
use a larger number of knots;  see \cite{Ruppert:02} and
\cite{Wang:11} for simulations and theory. Next we focus on
selecting the smoothing parameter.

We select the smoothing parameter by minimizing the pooled
generalized cross validation (PGCV), a functional extension of the
GCV (\citealt{Wahba:79}),
\begin{equation}
\label{gcv} \sum_{i=1}^I\left
\|\bY_i-\bS\bY_i\right\|^2/\{1-\tr(\bS)/J\}^2.
\end{equation}
Here $\|\cdot\|$ is the Euclidean norm of a vector.
Criterion~(\ref{gcv}) was also used in \cite{Zhang:07} and could be
interpreted as  smoothing each sample, $\bY_i$, using the same
smoothing parameter.  We argue that using criterion~(\ref{gcv}) is a
reasonable practice for covariance estimation. An alternative but
computationally hard method for selecting the smoothing parameter is
the leave-one-curve-out cross validation (\citealt{Yao:05}). The
following result indicates that PGCV can be easily calculated in
high dimensions.

\begin{prop}
\label{prop2} The PGCV in expression~(\ref{gcv}) equals to
$$
\frac{\sum_{k=1}^c C_{kk} (\lambda s_k)^2/(1+\lambda s_k)^2
-\|\bYtilde\|_F^2+ \|\bY\|_F^2}{\left\{1-J^{-1}\sum_{k=1}^c
(1+\lambda s_k)^{-1}\right\}^2},
$$
where $s_k$ is the $k$th element of $\bs$, $C_{kk}$ is the $k$th
diagonal element of $\bYtilde\bYtilde^T$, and $\|\cdot\|_F$ is the
Frobenius norm.
\end{prop}

The result shows that $\|\bY\|_F^2$, $\|\bYtilde\|_F^2$, and the
diagonal elements  of $\bYtilde\bYtilde^T$ need to be calculated
only once, which requires $O(IJ +cI)$ calculations. Thus, the FACE
algorithm is fast.

{\it FACE algorithm:}

{\it Step 1. Obtain the decomposition}
$(\mathbf{B}^T\mathbf{B})^{-1/2}\mathbf{P}(\mathbf{B}^T\mathbf{B})^{-1/2}
= \mathbf{U}{\rm diag}(\mathbf{s})\mathbf{U}^T$.

{\it Step 2. Specify  $\bS$ by  calculating and storing $\mathbf{s}$
and $\mathbf{A}_S =
\mathbf{B}(\mathbf{B}^T\mathbf{B})^{-1/2}\mathbf{U}$}.

{\it Step 3. Calculate and store $\bYtilde = \bA_S^T\bY$}.

{\it Step 4. Select $\lambda$ by minimizing PGCV in
expression~(\ref{gcv})}.

{\it Step 5. Calculate} $\bSigma_S = \left\{\mathbf{I}_{c}+\lambda
\textrm{diag}(\mathbf{s})\right\}^{-1}$.

{\it Step 6. Construct the decomposition} $
I^{-1}\bSigma_S\bYtilde\bYtilde^T\bSigma_S = \bA\bSigma\bA^T$.

{\it Step 7. Construct the decomposition} $\bKtilde =
(\bA_S\bA)\bSigma(\bA_S\bA)^T$.

The computation time of FACE  is $O\left(IJc +Jc^2+c^3 +
ck_0\right)$, where $k_0$ is the number of iterations needed for
selecting the smoothing parameter, and the total required memory  is
$O\left(IJ+I^2+Jc+c^2+k_0\right)$. See Proposition~\ref{prop3} in
the appendix for details. When $c = O(I)$ and $k_0 = o(IJ)$, the
computation time of FACE is $O(JI^2+I^3)$ and  $O(JI+I^2)$ memory
units are required. As a comparison, if we smooth the covariance
operator using other bivariate smoothers, then at least $O(J^2+IJ)$
memory units are required, which dramatically reduces the
computational efficiency of those smoothers.

\subsection{Estimating the scores}\label{sec::fpcs}
Under standard regularity conditions (\citealt{Karhunen:47}),
$X_i(t)$ can be written as $\sum_{k\geq 1} \xi_{ik} \psi_k(t)$ where
$\{\psi_k: k\geq 1\}$ is the set of eigenfunctions of $K$ and
$\xi_{ik} = \int_0^1 X_i(s)\psi_k(s)ds$ are the principals scores of
$X_i$. It follows that $ Y_i(t_j) = \sum_{k\geq 1}
\xi_{ik}\psi_k(t_j) +\epsilon_{ij}$. In practice,  we may be
interested in only the first $N$ eigenfunctions and approximate  $
Y_i(t_j)$ by  $\sum_{k=1}^N \xi_{ik} \psi_k(t_j) + \epsilon_{ij}$.
Using the estimated eigenfunctions $\psihat_k$'s and eigenvalues
$\lambdahat_k$'s from FACE, the scores of each $X_i$ can be obtained
by either numerical integration or as best  linear unbiased
predictors (BLUPs). FACE provides fast calculations of scores for
both approaches.

Let $\bYtilde_i$ denote the $i$th column of $\bYtilde$.  Let $\bxi_i
= (\xi_{i1},\dots, \xi_{iN})^T$ and let $\bAhat_N$ denote the first
$N$ columns of $\bA$ defined in Section~\ref{sec::spectral}. Let
$\bpsi_k = \{\psi_k(t_1),\dots, \psi_k(t_J)\}^T$ and $\bPsi =
[\bpsi_1,\dots, \bpsi_{N}]$.  The matrix $J^{-1/2}\bPsi$ is
estimated by $\bA_S\bAhat_N$. The method of numerical integration
estimates $\xi_{ik}$ by $ \xihat_{ik} = \int_0^1 Y_i(t) \psihat_k(t)
dt \approx J^{-1}\sum_{j=1}^J Y_i(t_j) \psihat_k (t_j). $


\begin{thm}
\label{thm1} The estimated principal scores $\bxihat_i =
(\xihat_{i1}, \dots, \xihat_{iN})^T$ using numerical integration are
$\bxihat_i = J^{-1/2} \bAhat_{N}^T\bYtilde_i, 1\leq i\leq I$.
\end{thm}


We now show how to obtain the estimated BLUPs for the scores. Let
$\epsilon_{ij} = Y_i(t_j) - \sum_{k=1}^{N} \psi_k(t_j)\xi_{ik}$ and
$\bepsilon_i = (\epsilon_{i1},\dots, \epsilon_{iJ})^T$. Then $ \bY_i
= \bPsi \bxi_i +\bepsilon_i$. The covariance  $\textrm{var}(\bxi_i)
= \text{diag}(\lambda_1,\dots, \lambda_{N})$ can be estimated by
$J\inv\bSigmahat_N = J\inv\text{diag}(\lambdahat_1,\dots,
\lambdahat_{N})$. The variance of $\epsilon_{ij}$ can be estimated
by
\begin{equation}
\label{sigma} \sigmahat^2=I\inv J\inv\|\bY\|_F^2-J^{-1}\sum_k
\hat{\lambda}_k.
\end{equation}

\begin{thm}
\label{thm2} Suppose  $\bPsi$ is estimated by
$J^{1/2}\bA_S\bAhat_N$, $\textrm{\textnormal{var}}(\bxi_i) =
\text{diag}(\lambda_1,\dots, \lambda_{N})$ is estimated by
$\bSigmahat_N = \text{diag}(\lambdahat_1,\dots, \lambdahat_{N})$,
and $\sigma^2$ is estimated by $\sigmahat^2$ in
equation~(\ref{sigma}). Then the estimated BLUPs of  $\bxi_i$ are
given by $ \bxihat_i = J^{-1/2}
\bSigmahat_{N}(\bSigmahat_{N}+J^{-1}\sigmahat^2\mathbf{I}_{N})^{-1}\bAhat_{N}^T\bYtilde_i$,
for $1\leq i\leq I$.
\end{thm}

Theorems~\ref{thm1} and~\ref{thm2} provide  fast approaches for
calculating the  principal scores using either numerical integration
or  BLUPs. These approaches combined with FACE are much faster
because they make use of the calculations already done for
estimating the eigenfunctions and eigenvalues. When $J$ is large,
the scores by BLUPs tend to be very close to those obtained by
numerical integration;  in the paper we only use numerical
integration.

\section{Extension of FACE}\label{sec::extension}
\subsection{Structured functional data}\label{sec::sfpca}
When analyzing structured functional data such as multilevel,
longitudinal, and crossed functional data (\citealt{Di:09,
Greven:10, Zipunnikov:11, Zipunnikov:12, Shou:13}), the covariance
matrices have been shown to be of the form  $\bY\bH\bY^T$, where
$\bH$ is a symmetric  matrix; see \cite{Shou:13} for more details.
We assume $\bH$ is  positive semi-definite because otherwise we can
replace $\bH$ by its positive counterpart.  Note that if $\bH_1$ is
a matrix such that  $\bH_1\bH_1^T = \bH$, smoothing $\bY\bH\bY^T$
can be done by using FACE for the transformed functional data
$\bY\bH_1$.  This insight is particularly useful for the sleep EEG
data, which has two visits and requires multilevel decomposition.

\subsection{Incomplete data}\label{sec::incomplete}
To handle incomplete data, such as the EEG sleep data where long
portions of the functions are unavailable, we propose an iterative
approach that alternates between covariance smoothing using FACE and
missing data prediction. Missing data are first initialized using a
smooth estimator of each individual curve within the range of the
observed data. Outside of the observed range the missing data are
estimated as the average of all observed values for that particular
curve. FACE is then applied to the initialized data, which produces
predictions of scores and functions and the procedure is then
iterated. We only use the scores of the first $N$ components, where
$N$ is selected by the criterion
\begin{equation*}
N = \min \left\{k: \frac{\sum_{j=1}^k \lambda_j}{\sum_{j=1}^{\infty}
\lambda_j}\geq 0.95\right\}.
\end{equation*}
Suppose $\hat{\boldsymbol{\Psi}}$ is the $p\times N$  matrix of
estimated eigenvectors  from FACE,
$\hat{\boldsymbol{\Sigma}}_N=\text{diag}(\hat\lambda_1,\dots,\hat\lambda_N)$
is the matrix of estimated eigenvalues, and
$\hat\sigma_{\epsilon}^2$ is the estimated variance of the noise.
Let $\mathbf{y}_{obs}$ denote the observed data  and
$\mathbf{y}_{mis}$ the missing data for a curve. Similarly,
$\hat{\boldsymbol{\Psi}}_{obs}$ is a sub-matrix of
$\hat{\boldsymbol{\Psi}}$ corresponding to the observed data  and
$\hat{\boldsymbol{\Psi}}_{mis}$ is another sub-matrix of
$\hat{\boldsymbol{\Psi}}$ corresponding to the missing data. Then
the prediction $(\hat{\mathbf{y}}_{mis},\hat{\boldsymbol{\xi}})$
minimizes the following
$$
\frac{\|\hat{\mathbf{y}}_{mis}-J^{1/2}\hat{\boldsymbol{\Psi}}_{mis}\hat{\boldsymbol{\xi}}\|_2^2
+
\|\mathbf{y}_{obs}-J^{1/2}\hat{\boldsymbol{\Psi}}_{obs}\hat{\boldsymbol{\xi}}\|_2^2}{2\hat\sigma_{\epsilon}^2}
+
\frac{1}{2}\hat{\boldsymbol{\xi}}^T\hat{\boldsymbol{\Sigma}}_N^{-1}
\hat{\boldsymbol{\xi}}.
$$
Note that if there is no missing data, the solution to this
minimization problem leads to Theorem~\ref{thm2}. For the next
iteration we replace $\mathbf{y}_{mis}$ by $\hat{\mathbf{y}}_{mis}$
and re-apply FACE to the updated complete data. We repeat the
procedure until convergence is reached.  In our experience
convergence is very fast and  typically achieved in fewer than $10$
iterations.

\section{The SSVD estimator and a subject-specific smoothing estimator}\label{sec::alternative}
A second approach for estimating the eigenfunctions  and eigenvalues
is to decompose the sample covariance matrix $\bKhat$ and then
smooth the eigenvectors. First let  $ \bU_y \bD_y \bV^T_y $ be the
singular value decomposition (SVD) of the data matrix $\bY$. Here
$\bU_y$ is a $J\times I$ matrix with orthonormal columns, $\bV_y$ is
an $I$ orthogonal matrix, and $\bD_y$ is an $I$ diagonal matrix. The
columns of $\bU_y$ contain all the eigenvectors of $\bKhat$ that are
associated with non-zero eigenvalues and the set of diagonal
elements of $I^{-1}\bD_y^2$ contain all the non-zero eigenvalues of
$\bKhat$. Thus, obtaining $\bU_y$ and
$\bD_y$ is equivalent to the eigendecomposition of $\bKhat$. 
Then we smooth the retained eigenvectors by smoothing splines,
implemented by the R function ``smooth.spline".   SSVD avoids the
direct decomposition of the sample covariance matrix and is
computationally simpler.  SSVD requires $O\{\min(I,J) IJ\}$
computations.

The approach of smoothing each curve and then diagonalizing the
sample covariance function of the smoothed curves can also be
efficiently implemented. First we smooth each curve using smoothing
splines. We  use the R function ``smooth.spline" which requires only
$O(J)$ computations for a curve with $J$ data points. Our experience
is that the widely used function ``gam" in the R package {\it mgcv}
(\citealt{Wood:13}) is much slower and can be computationally
intensive with a number of curves to smooth. Then instead of
directly diagonalizing the sample covariance of the smoothed curves,
which requires $O(J^3)$ computations, we calculate the singular
value decomposition of the $I\times J$ matrix formed by the smoothed
curves, which requires only $O(\min(I,J)IJ)$ computations. The
resulting right singular vectors estimate the  eigenfunctions scaled
by $J^{-1/2}$. Without the SVD step, a brute-force decomposition of
the $J\times J$ sample covariance becomes infeasible when $J$ is
large, such as $5,000$.  We will refer to the this approach as
S-Smooth, which, to the best of our knowledge, is the first
computationally efficient method  for covariance estimation using
subject-specific smoothing.

We will compare SSVD, S-Smooth and FACE in terms of performance and
computation time in the simulation study.




\section{Simulation}\label{sec::sim}

We consider three simulation studies. In the first study we use
moderately high-dimensional data contaminated with noise. We let
$J=3,000$ and $I=50$, which are roughly the dimensions of the EEG
data in Section~\ref{sec::example}.  We use SSVD, S-Smooth and FACE.
We did not evaluate other  bivariate smoothers  because we were
unable to run them on such dimensions in a reasonably short time. In
the second  study we consider functional data where portions of the
observed functions are missing completely at random (MCAR). This
simulation is directly inspired by our EEG data where long portions
of the functions are missing. In the last  study we assess the
computation time of FACE and compare it with that of SSVD and
S-Smooth. We also provide the computation time of the sandwich
smoother (\citealp{Xiao:12}).  We use R code that is made available
with this paper. All simulations are run on modest, widely available
computational resources: an Intel Core i5 2.4 GHz Mac with 8
gigabytes  of random access memory.

\subsection{Complete data}\label{sec::cov_sim}
We consider the following covariance functions:
\begin{enumerate}
\item[1\&2]  {\bfseries Finite basis expansion}. $K(s,t) = \sum_{\ell=1}^3 \lambda_{\ell}\psi_{\ell}(s)\psi_{\ell}(t)$ where $\psi_{\ell}$'s are eigenfunctions and  $\lambda_{\ell}$'s are eigenvalues. We choose $\lambda_{\ell} = 0.5^{\ell-1}$ for $\ell=1,2,3$ and there are two sets of eigenfunctions: case 1: $\psi_1(t) = \sqrt{2}\sin(2\pi t)$, $\psi_2(t) = \sqrt{2}\cos(4\pi t)$ and $\psi_3(t) = \sqrt{2}\sin(4\pi t)$; and case 2: $\psi_1(t) = \sqrt{3}(2t-1)$, $\psi_2(t) = \sqrt{5}(6t^2-6t+1)$ and $\psi_3(t) = \sqrt{7}(20t^3-30t^2+12t-1)$.

\item[3] {\bfseries Brownian motion}. $K(s,t) = \sum_{\ell=1}^{\infty} \lambda_{\ell}\psi_{\ell}(s)\psi_{\ell}(t)$ with  eigenvalues $\lambda_{\ell} = \frac{1}{(\ell-1/2)^2\pi^2}$ and eigenfunctions $\psi_{\ell}(t) = \sqrt{2}\sin((\ell-1/2)\pi t)$.

\item[4] {\bfseries Brownian bridge}. $K(s,t) = \sum_{\ell=1}^{\infty} \lambda_{\ell}\psi_{\ell}(s)\psi_{\ell}(t)$ with eigenvalues $\lambda_{\ell} = \frac{1}{\ell^2\pi^2}$ and eigenfunctions $\psi_{\ell}(t) = \sqrt{2}\sin(\ell \pi t)$.

\item[5] {\bfseries Mat$\mathbf{\acute{\textbf{e}}}$rn covariance structure}. The Mat$\acute{\textnormal{e}}$rn covariance function
$$
C(d;\phi,\nu) =
\frac{1}{2^{\nu-1}\Gamma(\nu)}\left(\frac{\sqrt{2\nu}d}{\phi}\right)^{\nu}
K_{\nu}\left(\frac{\sqrt{2\nu}d}{\phi}\right)
$$
with  range $\phi=0.07$ and order $\nu=1$. Here $K_{\nu}$ is the
modified Bessel function of order $\nu$. The top three eigenvalues
for this covariance function are $0.209$, $0.179$ and $0.143$.
\end{enumerate}

We generate data at $\{1/J,2/J,\dots, 1\}$ with $J=3,000$ and add
i.i.d.\ $ \mathcal{N}(0,\sigma^2)$ errors to the data. We let
\[
\sigma^2 = \int_{s =0}^1\int_{t=0}^1 K(s,t) \mathrm{d}s\mathrm{d}t,
\]
which implies that  the signal to noise ratio in the data is $1$.
The number of curves is $I = 50$ and for each covariance function
200 datasets are drawn.

\begin{figure*}[htp]
\begin{center}
\includegraphics[width=6.5in, angle=0]{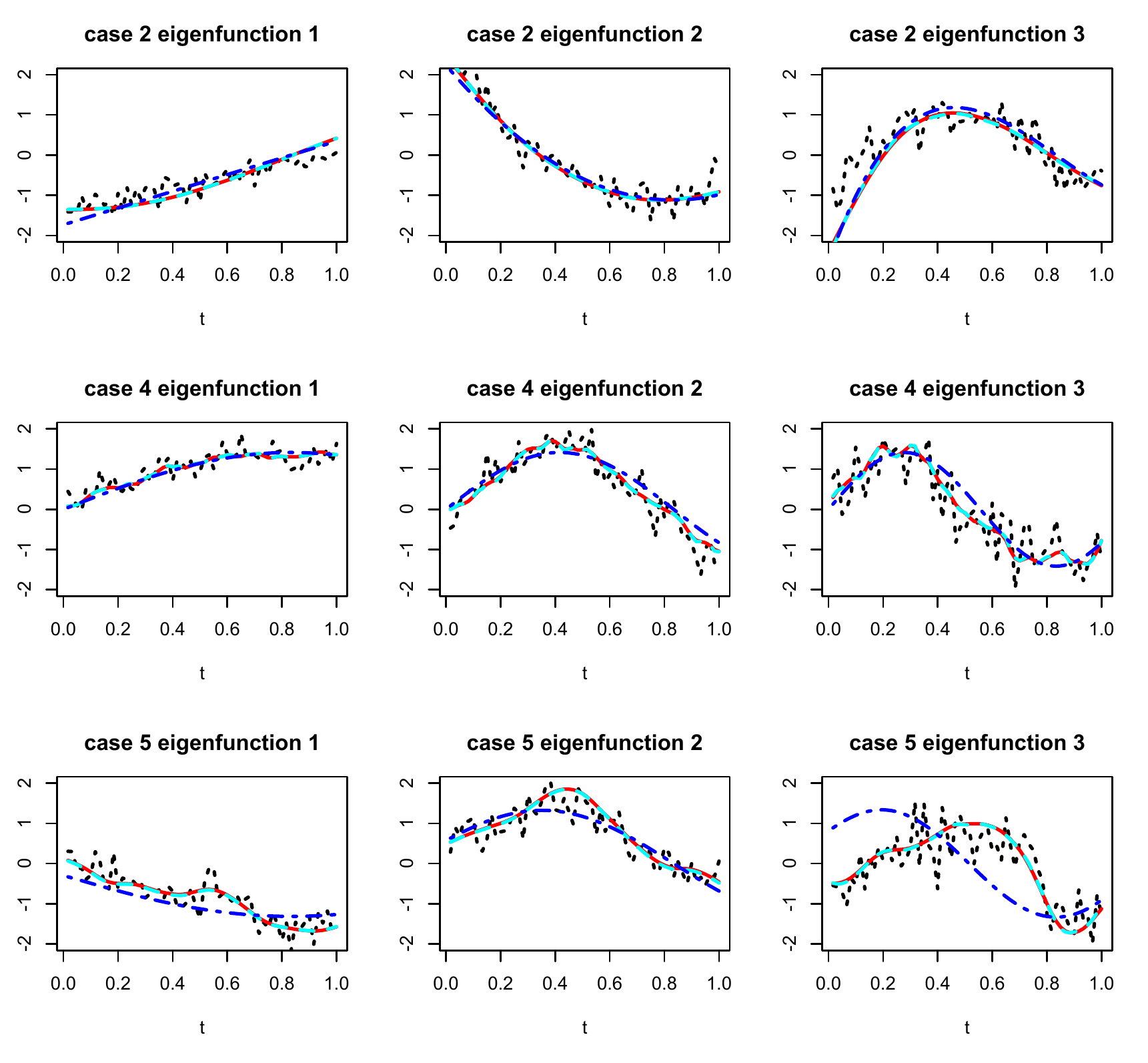}
\end{center}
\caption{\label{fig_est_eigenfunction}True and estimated
eigenfunctions for three cases each with one simulated data set.
Each row corresponds to one simulated data set.  Each box shows the
true eigenfunction (blue dot-dashed lines), the estimated
eigenfunction using FACE (red solid lines),  the estimated
eigenfunction using SSVD (cyan dashed lines), and the estimated
eigenfunction without smoothing (black dotted lines). We do not show
the estimates from S-Smooth and FACE (incomplete data) because they
are almost identical to these from FACE and SSVD.}
\end{figure*}

We compare the performance of the three methods to estimate: (1) the
covariance matrix; (2)  the eigenfunctions; and (3) the eigenvalues.
For simplicity, we only consider the top three
eigenvalues/eigenfunctions. For FACE we use 100 knots; for SSVD and
S-Smooth we use smoothing splines, implemented through the R
function `smooth.spline'. Figure~\ref{fig_est_eigenfunction}
displays, for one simulated data set for each case,  the true and
estimated eigenfunctions using SSVD and FACE, as well as the
estimated eigenfunctions without smoothing.

We see from Figure~\ref{fig_est_eigenfunction} that the smoothed
eigenfunctions  are very similar and  the estimated eigenfunctions
without smoothing are quite noisy. The results are expected as all
smoothing-based methods are designed to account for the noise in the
data and the discrepancy between the estimated and
the true eigenfunctions  is mainly due to the variation in the random
functions. Table~\ref{table_mise_eigenfunction} provides the mean
integrated squared errors (MISE) of the estimated eigenfunctions
indicating that FACE and S-Smooth have  better performance than
SSVD. For case 5, the smoothed eigenfunctions for all methods are
far from the true eigenfunctions. This is not surprising because for
this case the eigenvalues are close to each other and it is known
that the accuracy of eigenfunction estimation also depends on the
gap between consecutive eigenvalues; see for example,
\cite{Bunea:13}.
 In terms of covariance estimation, Table~\ref{table_mise_covariance}
 suggests that SSVD is outperformed by the other two methods. However,
 the simplicity and robustness of SSVD may actually make it quite
 popular in applications.

\begin{figure*}
\centering
\includegraphics[width = 6.5in, angle=0]{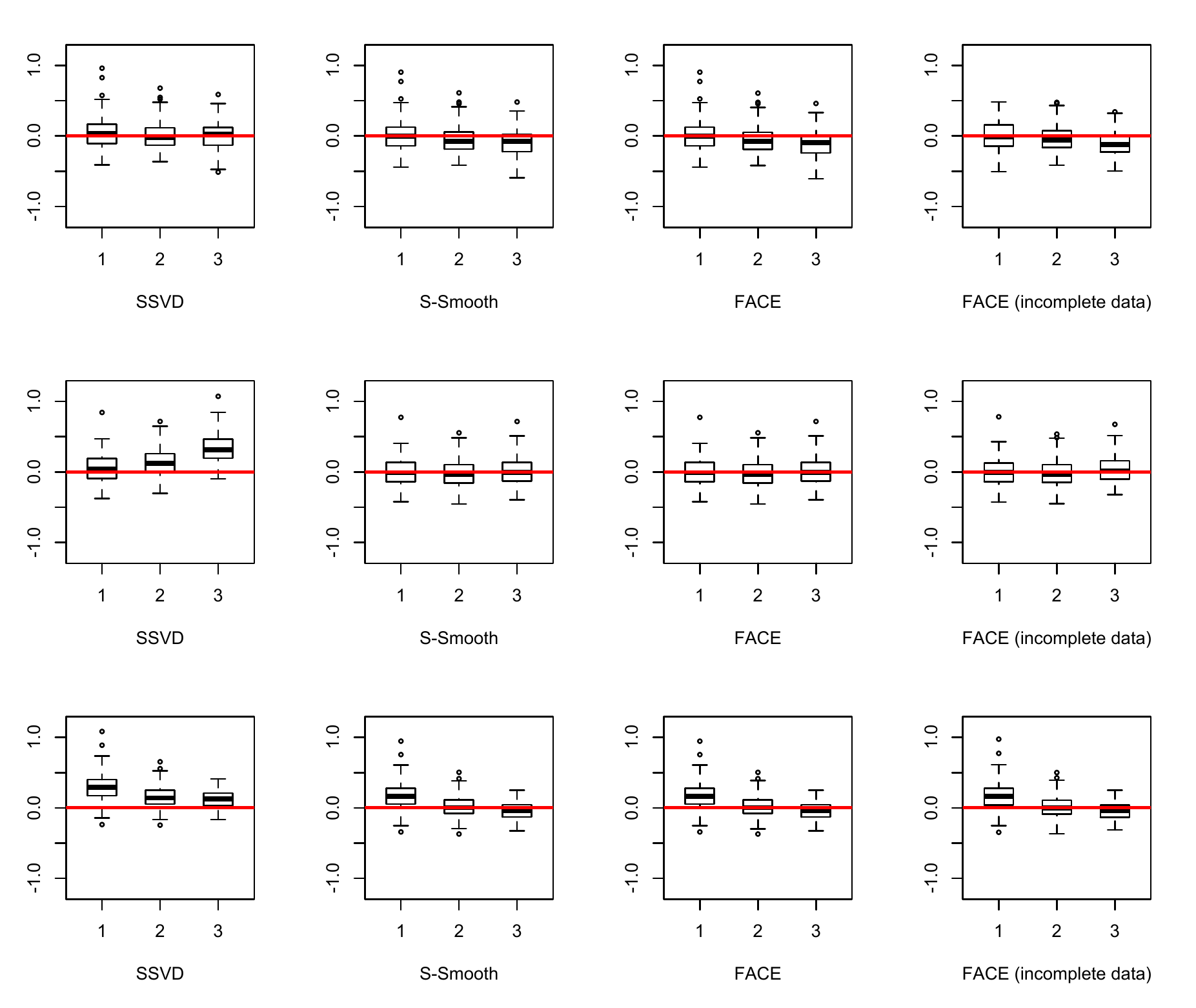}
\caption{\label{fig.eigen2}Boxplots of the centered and standardized
estimated eigenvalues, $\lambdahat_k/\lambda_k-1$. The top panel is
for case 2, the middle panel is for case 4,  and the bottom
panel is for case 5. The zero is shown by the solid red line. Case
1 is similar to case 2 and case 3 is similar to case 4,
and hence are not shown. }
\end{figure*}

Figure~\ref{fig.eigen2} shows boxplots of estimated eigenvalues that
are centered and  standardized, $\lambdahat_k/\lambda_k-1$.  The
SSVD method works well for cases 1 and 2, where the true
covariance has only three non-zero eigenvalues, but tends to
overestimate the eigenvalues  for the other three cases,  where the
covariance function has an infinite number of non-zero eigenvalues.
In contrast, the  FACE and S-Smooth estimators underestimate the
eigenvalues for the simple cases 1 and 3 but  are much closer
to the true eigenvalues for the more complex cases.
Table~\ref{table_mise_eigenvalue}  provides the average mean squared
errors (AMSEs) of $\hat\lambda_k/\lambda_k - 1$ for $k=1, 2, 3$, and
indicates that S-Smooth and FACE tend to estimate the eigenvalues
more accurately.

\subsection{Incomplete data}
In Section~\ref{sec::incomplete} we extended FACE for incomplete
data, and here we  illustrate the extension with a simulation. We
use the same simulation setting in Section~\ref{sec::cov_sim} except
that for each subject we allow for portions of observations missing
completely at random. For simplicity we fix the length of each
portion so that $0.065J$ consecutive observations are missing. We
allow one subject to miss either 1, 2, or 3 portions with equal
probabilities so that in expectation $13\%$ of the data are missing.
Note that the real data we will consider later also has about $13\%$
measurements missing.

In Figure~\ref{fig.eigen2}, boxplots of the estimated eigenvalues
are  shown.  The MISEs of the  estimated covariance function and
estimated eigenfunctions and the AMSEs of the estimated eigenvalues
appear in Tables~\ref{table_mise_covariance},
\ref{table_mise_eigenfunction} and~\ref{table_mise_eigenvalue},
respectively.  The simulation results show that the performance of
FACE degrades only marginally.

\subsection{Computation time}
We record the computation time of FACE for various combinations of
$J$ and $I$.  All other settings remain the same as in the first
simulation study and we use the eigenfunctions from case 1.  For
comparison the computation times of SSVD, S-Smooth and the sandwich
smoother (\citealt{Xiao:12}) are also given. Table~\ref{time}
summarizes the results and shows that FACE is fast even with
high-dimensional data while the computation time of the sandwich
smoother increases dramatically with $J$, the dimension of the
problem. For example it took FACE only 5 seconds to smooth a 10,000
by 10,000 dimensional matrix for $500$ subjects, while the sandwich
smoother did not run on our computer. While SSVD, S-Smooth and FACE
are all fast to compute,  FACE is computationally faster when
$I=500$. We note that S-Smooth has additional problems when data are
missing, though a method similar to FACE may be devised. Ultimately,
we prefer the self-contained, fast, and flexible FACE approach.

Although we do not run FACE on ultrahigh-dimensional data,  we can
obtain a rough estimate of  the computation time by the formula
$O(JIc)$. Table~\ref{time} shows that FACE with 500 knots takes 5
seconds on data with $(J,I) = (10000, 500)$. For data with $J$ equal
to 100,000 and $I$ equal to 2,000, FACE with 500 knots should take 4
minutes to compute, without taking into account the time for loading
data into the computer memory. Our code was written and run in R, so
a faster implementation of FACE may be possible on other software
platforms.


\section{Example}\label{sec::example}
\begin{figure*}
\centering
\includegraphics[width=6in,angle=0]{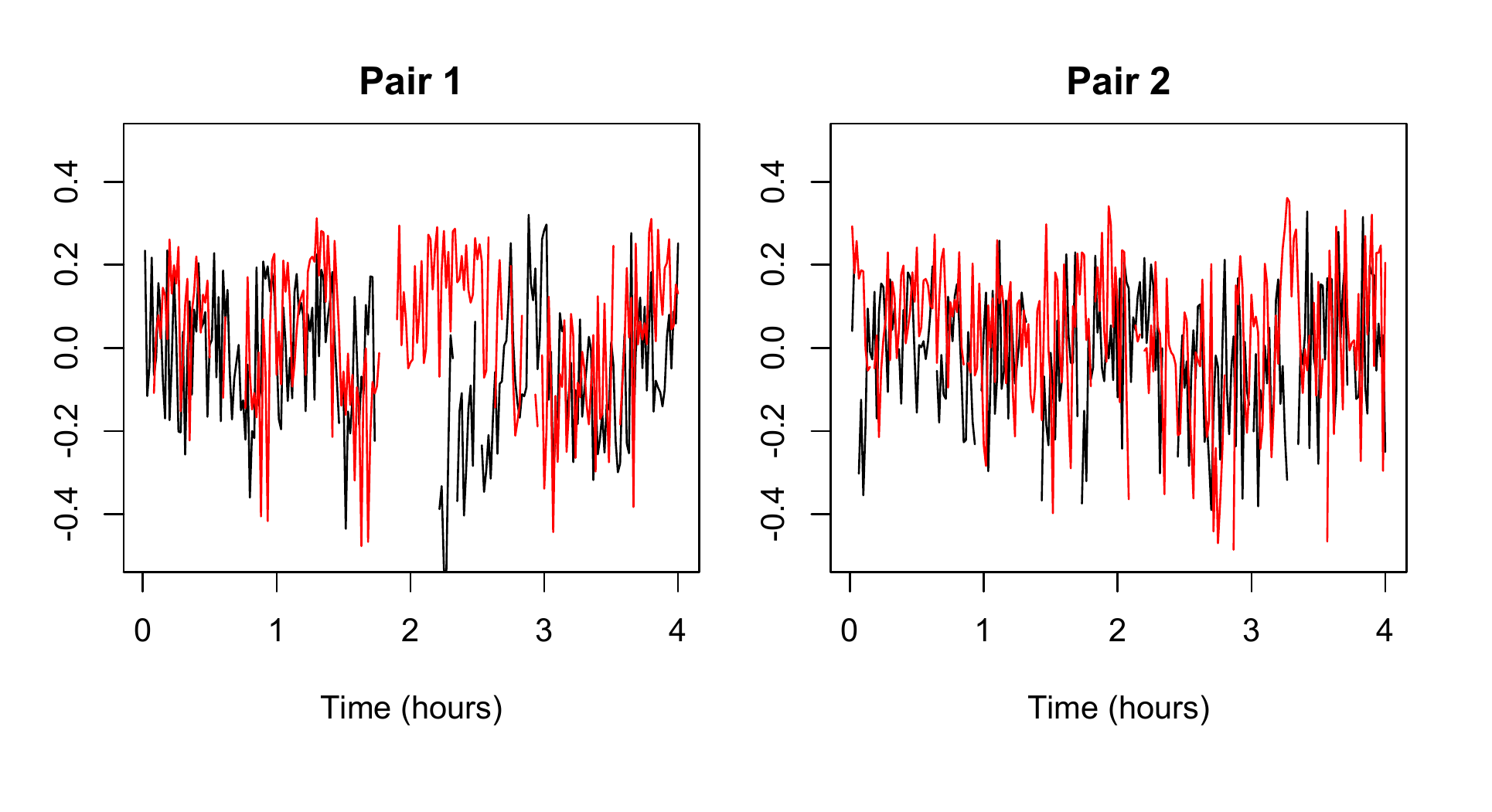}
\caption{\label{fig.data} Data for two matched pairs of case and
controls in the Sleep Heart Health Study. The red lines are for
cases while the black are for controls. For simplicity only the last
observation in each minute of the 4-hour interval is shown. }
\end{figure*}

The Sleep Heart Health Study (SHHS) is a large-scale study of sleep
and its association with health-related outcomes. Thousands of
subjects enrolled in SHHS underwent two in-home polysomnograms
(PSGs) at multiple visits. Two-channel electroencephalographs (EEG),
part of the PSG,  were collected at a frequency of 125Hz, or 125
observations per second for each subject, visit and channel. We
model the proportion of $\delta$-power which is a summary measure of
the spectrum of the EEG signal. More details on $\delta$-power can
be found in \cite{Crainiceanu:09} and \cite{Di:09}.   The data
contain   51 subjects with sleep-disordered breathing (SDB) and 51
matched controls; see \cite{Crainiceanu:12} and \cite{Swihart:12}
for details on how the pairs were matched. An important feature of
the EEG data is that  long consecutive portions of observations,
which indicate wake periods, are missing. Figure~\ref{fig.data}
displays data from 2 matched pairs. In total about $13\%$ of the
data is missing.

Similar to \cite{Crainiceanu:12}, we consider the following
statistical model. The data for proportion of $\delta$-power  are
pairs of curves $\{Y_{iA}(t), Y_{iC}(t)\}$, where $i$ denotes
subject, $t=t_1, \dots, t_J\ (J = 2,880)$ denotes the time measured
in 5-second intervals in a 4-hour sleep interval from sleep onset,
$A$ stands for apneic and $C$ stands for control.  The model is \beq
\label{example.model} \left\{
\begin{array}{l}
Y_{iA}(t) = \mu_A(t) + X_i(t) +  U_{iA}(t) + \epsilon_{iA}(t)\\
Y_{iC}(t) = \mu_C(t) + X_i(t) + U_{iC}(t) + \epsilon_{iC}(t)\\
\end{array}
\right. \eeq
where $\mu_A(t)$ and $\mu_C(t)$ are  mean functions of  proportions of $\delta$-power, $X_i(t)$ is a functional process with mean 0 and  continuous covariance operator $K_X(\cdot, \cdot)$, $U_{iA}(t)$ and $U_{iC}(t)$ are functional processes with mean 0 and continuous  covariance operator $K_U(\cdot,\cdot)$, and $\epsilon_{iA}(t), \epsilon_{iC}(t)$ are measurement errors with mean 0 and variance $\sigma^2$. The random processes $X_i, U_{iA}, U_{iC}, \epsilon_{iA}$ and $\epsilon_{iC}$ are assumed to be mutually independent. Here $X_i$ accounts for the between-pair correlation of the data while $U_{iA}$ and $U_{iC}$ model the within-pair correlation. The Multilevel Functional Principal Component Analysis (MFPCA) (\citealt{Di:09}) can be used to analyze data with model~(\ref{example.model}). One crucial step of MFPCA is to smooth  two estimated covariance operators which in this example are $2880\times 2880$ matrices. 

\begin{figure*}
\centering
\includegraphics[angle=0]{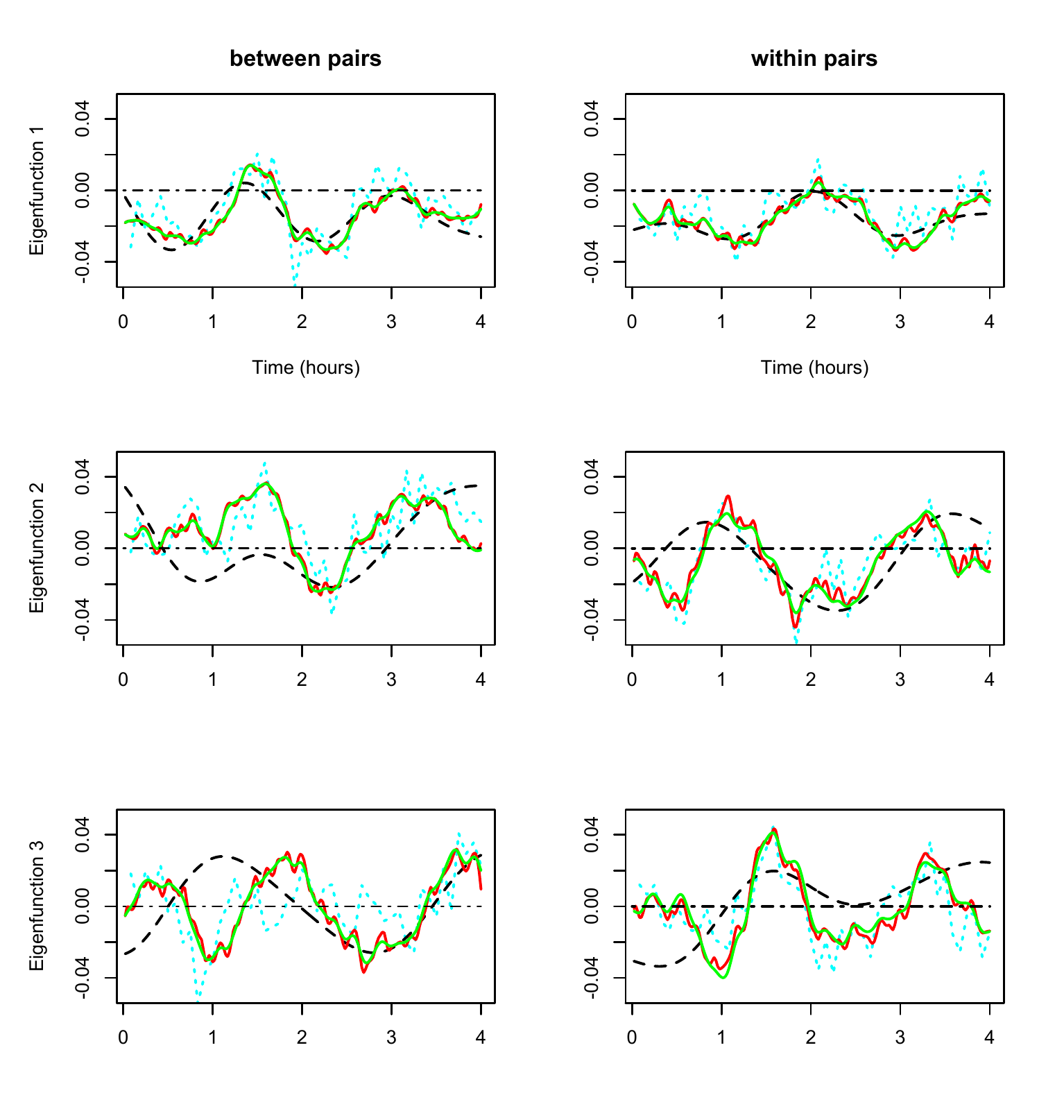}
\caption{\label{fig.example} The eigenfunctions associated with the
top three eigenvalues of  $K_X$ and $K_U$ for the Sleep Heart Health Study data. The left column is for
$K_X$ and the right one is for $K_U$.  The red and green solid lines
correspond to the FACE approach using the original and modified GCV,
respectively. The black dashed lines are for thin plate splines, and
the cyan dotted lines are for SSVD.}
\end{figure*}

Smoothing  large covariance operators of dimension $2880\times 2880$
can be computationally expensive. We tried bivariate thin plate
regression splines  and used  the R function `{\it bam}' in the {\it
mgcv} package (\citealt{Wood:13}) with 35 equally-spaced knots for
each axis. The smoothing parameter was automatically  selected by
`{\it bam}' with the option `GCV.cp'. Running time for thin plate
regression splines was three hours. Because the two covariance
operators take the form in Section~\ref{sec::sfpca} (see the details
in Appendix~\ref{appendix::H}), we applied FACE, which ran in less
than 10 seconds with 100 knots. Note that we also tried thin plate
splines with 100 knots in {\it mgcv}, which was still running after
10 hours.  Figure~\ref{fig.example} displays the first three
eigenfunctions for $K_X$ and $K_U$, using both methods. As a
comparison,
the eigenfunctions using SSVD are also shown. For the SSVD method, to handle incomplete data the SVD step was replaced by a brute-force decomposition of the two $2880\times 2880$ covariance operators. Figure~\ref{fig.example} shows that the top eigenfunctions obtained from the two bivariate smoothing methods are quite different, except for the first eigenfunctions on the top row. The estimated eigenfunctions using FACE in general resemble those by SSVD with some subtle differences, while thin plate splines in this example seem to over-smooth the data, probably because we were forced to use a smaller number of knots. 

The smoothed eigenfunctions from FACE using PGCV (red solid lines in
Figure~\ref{fig.example}) appear undersmooth.
 This may be due to the well
reported tendency of GCV to undersmooth as well as to the
noisy and complex nature of the data. 
A common way to combat this problem is to use modified GCV (modified
PGCV for our case)  where  $\tr(\bS)$ in~(\ref{gcv}) is multiplied
by a constant $\alpha$ that is greater than 1; see \cite{Cummins:01}
and \cite{Kim:04} for such practices for smoothing splines.  Similar
practice has also been proposed for AIC in \cite{Shinohara:14}.  We
re-ran the FACE method with $\alpha = 2$ and the resulting estimates
(green solid lines in Figure~\ref{fig.example}) appear  more
satisfactory.  In this case, the direct smoothing approach of the
eigenfunctions (\citealt{Rice:91,Capra:97,Ramsay:05}) might provide
good results. However, the missing data issue and the computational
difficulty associated with large $J$ make the approach difficult to
use.

Table~\ref{eigenvalues} provides  estimated eigenvalues of $K_X$ and
$K_U$.  Compared to FACE (with $\alpha=2$), thin plate splines
over-shrink significantly the eigenvalues, especially those of the
between pair covariance.  The results from FACE in
Table~\ref{eigenvalues} show that the proportion of variability
explained by $K_X$, the between-pair variation, is
$14.40/(14.40+22.75) \approx 38.8\%$.

\section{Discussion}\label{sec::dis}
In this paper  we developed a fast covariance estimation (FACE)
method  that could significantly alleviate the computational
difficulty of bivariate smoothing and eigendecomposition of large
covariance matrices in  FPCA for high-dimensional data. Because
bivariate smoothing and eigendecomposition of covariance matrices
are integral parts of FPCA, our method could increase the scope and
applicability of FPCA for high-dimensional data. For instance, with
FACE, one may consider  incorporating high-dimensional functional
predictors into  the penalized functional regression model of
\cite{Goldsmith:11}.

The proposed FACE method can be regarded as a two-step procedure
such as S-Smooth (see, e.g., \citealt{Besse:86, Ramsay:91, Besse:97,
Cardot:00, Zhang:07}). Indeed, if we first smooth
 data at the subject level $\bYhat_i = \bS\bY_i, i=1,\dots, I$,
then it is easy to show that the empirical covariance estimator of
the $\bYhat_i$ is equal to $\widetilde{\bK}$. There are, however,
important computational differences between FACE and the current
two-step procedures. First, the fast algorithm in
Section~\ref{sec::gcv} enables FACE to select efficiently the
smoothing parameter. Second, FACE could work with structured
functional data and allow for different smoothing for each
covariance operator. Third, FACE can be easily extended for
incomplete data where long consecutive portions of data are missing
while it is unclear how a two-step procedure could be used for such
data.

The second approach, SSVD, is very simple and reasonable, though
some problems remain open, especially in applications with missing
data. Another drawback of SSVD is that the smoothed eigenvectors are
not necessarily orthogonal, though the fast Gram-Schmidt algorithm
could easily be applied to the smooth vectors. Overall, we found
that using a combination of FACE and SSVD provides a reasonable and
practical starting point for smoothing covariance operators for high
dimensional functional data, structured or unstructured.

In this paper we have only considered the case when the sampling
points are the same for all subjects. Assume now for the $i$th
sample that we observe $\bY_i = \{Y_i(t_{i1}),\dots,
Y_i(t_{iJ_i})\}^T$, where $t_{ij}$, $j=1,\ldots,J_i$ can be
different across subjects. In this case the empirical estimator of
the covariance operator does not have a decomposable form. Consider
the scenario when subjects are densely sampled and all $J_i$'s are
large. Using the idea from \cite{Di:09}, we can undersmooth each
$\bY_i$ using, for example, a kernel smoother with a small bandwidth
or a regression spline. FACE can then be applied on the
under-smoothed estimates
 evaluated at  an equally spaced grid, $\{\bYhat_1,\dots,\bYhat_I\}$. Extension of FACE to the sparse design scenario remains  a difficult open problem.

\section*{Acknowledgement}
This work was supported by Grant Number R01EB012547 from the
National Institute of Biomedical Imaging And Bioengineering and
Grant Number R01NS060910 from the National Institute of Neurological
Disorders and Stroke. This work represents the opinions of the
researchers and not necessarily that of the granting organizations.

\appendix
\section{Appendix: Proofs}

{\it Proof of Proposition~\ref{prop1}}: The design matrix $\bB$ is
of full rank (\citealt{Xiao:11}). Hence $\bB^T\bB$ is invertible and
$\bA_S$ is of rank $c$. $\bSigma_S$ is a diagonal matrix with all
elements greater than 0 and $\bYtilde$ is of rank at most
$\min(c,I)$. Hence $\bKtilde =
\bA_S\left(I^{-1}\bSigma_S\bYtilde\bYtilde^T\bSigma_S\right)\bA_S^T$
has a rank at most $\min(c,I)$ and the proposition follows.

{\it Proof of Proposition~\ref{prop2}}: First of all,
$\tr(\mathbf{S}) = \tr(\bSigma_S)$ which  is easy to calculate.  We
now compute $\sum_{i=1}^I \|\bY_i - \bS\bY_i\|^2$. Because $ \|\bY_i
- \bS\bY_i\|^2 =  \bY_i^T(\bS-\bI_J)^2\bY_i = \tr
\{(\bS-\bI_J)^2\bY_i\bY_i^T\}$,
\[
\sum_{i=1}^I \|\bY_i - \bS\bY_i\|^2 =
\tr\left\{(\bS-\bI_J)^2\sum_{i=1}^I \bY_i\bY_i^T\right\} =
\tr\left\{(\bS-\bI_J)^2\bY\bY^T\right\}.
\]
It can be shown that $\bS^2 = \bA_S \bSigma_S^2\bA_S^T$. Hence
$\tr(\bS^2\bY\bY^T) = \tr(\bY^T\bS^2\bY) =
\tr(\bYtilde^T\bSigma_S^2\bYtilde)=\tr(\bSigma_S^2\bYtilde\bYtilde^T)$.
Similarly, we derive $\tr(\bS\bY\bY^T) =
\tr(\bSigma_S\bYtilde\bYtilde^T)$.  We have $\tr(\bY\bY^T) =
\|\bY\|_F^2$. It follows that
\begin{equation*}
\sum_{i=1}^I \|\bY_i - \bS\bY_i\|^2 =
\tr\left\{(\bSigma_S-\bI_c)^2\bYtilde\bYtilde^T\right\}
-\|\bYtilde\|_F^2+ \|\bY\|_F^2.
\end{equation*}

\begin{prop}
\label{prop3} The computation time of FACE  is $O\left(IJc +Jc^2+c^3
+ ck_0\right)$, where $k_0$ is the number of iterations needed for
selecting the smoothing parameter (see Section~\ref{sec::gcv}), and
the total required computer memory  is
$O\left(JI+I^2+Jc+c^2+k_0\right)$ memory units.
\end{prop}
{\it Proof of Proposition~\ref{prop3}}:  We need to compute or store
the following quantities: $\bX$, $\bB$,  $\bB^T\bB$,
$(\bB^T\bB)^{-1/2}$,  $\bP$,
$(\bB^T\bB)^{-1/2}\bP(\bB^T\bB)^{-1/2}$, $\bA_S$, $\bYtilde$,
$\bA$, $\bU$, and $\bA_S \bA$. For the computational complexity,
$\bB^T\bB$, $\bA_S = \bB (\bB^T\bB)^{-1/2}\bU$, and $\bA_S\bA$
require $O(Jc^2)$ computations; $(\bB^T\bB)^{-1/2}$, $\bP$,
$(\bB^T\bB)^{-1/2}\bP(\bB^T\bB)^{-1/2}$, $\bA$, and  $\bU$ require
$O(c^3)$ computations; $\bYtilde = \bA_S^T\bY$ requires $O(JIc)$
computations. So in total, $O(JIc+Jc^2+c^3)$ computations are
required. For the memory burden, the loading of $\bY$ requires
$O(JI)$ memory units, computer of $\bB$ and $\bA_S\bA$ requires
$O(Jc)$ memory units, and other objects require $O(c^2)$ memory
units.

{\it Proof of Theorem~\ref{thm1}}:  We have $\bxihat_i =
J^{-1/2}(\bA_S\bAhat_N)^T\bY_i = J^{-1/2}\bAhat_N^T
(\bA_S^T\bY_i)=J^{-1/2}\bAhat_N^T\bYtilde_i$.
{\it Proof of Theorem~\ref{thm2}}:
 Let $\bAtilde_N$ denote the first $N$ columns of $\bA_S\bA$, then $\bAtilde_N = \bA_S\bAhat$.
The estimated BLUPs for $\bxi_i$ (\citealt{Ruppert:03}) is
$$
\bxihat_i = J^{-1/2}\bSigmahat_N\bAtilde_N^T
\left(\bAtilde_N\bSigmahat_N\bAtilde_N^T +J\inv
\sigmahat^2\bI_J\right)^{-1}\bY_i.
$$
The inverse matrix in the above equality can be replaced by the
following (\cite{Seber:07}, page 309, equality b(i)),
\[
 \left(\bAhat_N\bSigmahat_N\bAtilde_N^T + J\inv \sigmahat^2\bI_J\right)^{-1} = \frac{J}{\sigmahat^2}\left\{\bI_N- \frac{J}{\sigmahat^2} \bAtilde_N\left( \bSigmahat_N^{-1} +\frac{J}{\sigmahat^2}\bI_N\right)^{-1} \bAtilde_N^T\right\}.
\]
It follows that \beqn
\bxihat &=&  J^{-1/2}\frac{J}{\sigmahat^2}\bSigmahat\left\{\bI_N -  \frac{J}{\sigmahat^2}\left( \bSigmahat_N^{-1} +\frac{J}{\sigmahat^2}\bI_N\right)^{-1}\right \}\bAhat_N^T\bYtilde_i\\
& =& J^{-1/2} \bSigmahat_N\left(\bSigmahat_N +
J^{-1}\sigmahat^2\bI_N\right)^{-1}\bAhat_N^T\bYtilde_i. \eeqn

\section{Appendix: Empirical covariance operators for $K_X$ and $K_U$}\label{appendix::H}
Let $I$ denote the number of pairs of cases and controls. For
simplicity, we assume estimates of $\mu_A(t)$ and $\mu_C(t)$ have
been subtracted from $Y_{iA}$ and $Y_{iC}$, respectively. Let
$\bY_{iA} = (Y_{iA}(t_1),\dots, Y_{iA}(t_T))^T$ and $\bY_{iC} =
(Y_{iC}(t_1),\dots, Y_{iC}(t_J))^T$. By \cite{Zipunnikov:11}, we
have estimates of the covariance operators,
\[
\bKhat_X = \frac{1}{2I}\sum_{i=1}^I\left(\bY_{iA}\bY_{iC}^T +
\bY_{iC}\bY_{iA}^T\right),
\]
and
\[
\bKhat_U =
\frac{1}{2I}\sum_{i=1}^I\left(\bY_{iA}-\bY_{iC}\right)\left(\bY_{iA}-\bY_{iC}\right)^T.
\]
Let $\bY_A = [\bY_{1A},\dots, \bY_{nA}]$, $\bY_C = [\bY_{1C}, \dots,
\bY_{nC}]$ and $\bY = [\bY_A,\bY_C]$. Then $\bY$ is of dimension
$J\times 2I$.  It can be shown that $\bKhat_X = \bY\bH_X\bY^T$ and
$\bKhat_U = \bY\bH_U\bY^T$, where
\[
\bH_X = \frac{1}{2I}\left(\begin{array}{cc} \mathbf{0}_{I}& \bI_{I}\\
\bI_{I}&\mathbf{0}_{I}
\end{array}\right),\,\,
\bH_U = \frac{1}{2I}\left(\begin{array}{cc} \mathbf{I}_{I}& -\bI_{I}\\
-\bI_{I}&\mathbf{I}_{I}
\end{array}\right).
\]

\bibliographystyle{chicago}
\bibliography{facebib}

\begin{thebibliography}{}

\bibitem[\protect\citeauthoryear{Besse, Cardot, and Ferraty}{Besse
  et~al.}{1997}]{Besse:97}
Besse, P., H.~Cardot, and F.~Ferraty (1997).
\newblock {Simultaneous nonparametric regressions of unbalanced longitudinal
  data}.
\newblock {\em Comput. Statist. Data Anal.\/}~{\em 24}, 255--270.

\bibitem[\protect\citeauthoryear{Besse and Ramsay}{Besse and
  Ramsay}{1986}]{Besse:86}
Besse, P. and J.~O. Ramsay (1986).
\newblock {Principal components analysis of sampled functions}.
\newblock {\em Psychometrika\/}~{\em 51}, 285--311.

\bibitem[\protect\citeauthoryear{Bunea and Xiao}{Bunea and
  Xiao}{2013}]{Bunea:13}
Bunea, F. and L.~Xiao (2013).
\newblock {On the sample covariance matrix estimator of reduced effective rank
  population matrices, with applications to fPCA}.
\newblock To appear in {\it Bernoulli}, available at
  \url{http://arxiv.org/abs/1212.5321}.

\bibitem[\protect\citeauthoryear{Capra and M\"uller}{Capra and
  M\"uller}{1997}]{Capra:97}
Capra, W. and H.~M\"uller (1997).
\newblock {An accelerated-time model for response curves}.
\newblock {\em J. Amer. Statist. Assoc.\/}~{\em 92}, 72--83.

\bibitem[\protect\citeauthoryear{Cardot}{Cardot}{2000}]{Cardot:00}
Cardot, H. (2000).
\newblock {Nonparametric estimation of smoothed principal components analysis
  of sampled noisy functions}.
\newblock {\em J. Nonparametr. Statist.\/}~{\em 12}, 503--538.

\bibitem[\protect\citeauthoryear{Crainiceanu, Reiss, Goldsmith, Huang, Huo,
  Scheipl, Swihart, Greven, Harezlak, Kundu, Zhao, Mclean, and
  Xiao}{Crainiceanu et~al.}{2013}]{Crainiceanu:13}
Crainiceanu, C., P.~Reiss, J.~Goldsmith, L.~Huang, L.~Huo, F.~Scheipl,
  B.~Swihart, S.~Greven, J.~Harezlak, M.~Kundu, Y.~Zhao, M.~Mclean, and L.~Xiao
  (2013).
\newblock {R package {\it refund}: Methodology for regression with functional
  data (version 0.1-9)}.
\newblock URL:\url{http://cran.r-project.org/web/packages/refund/index.html}.

\bibitem[\protect\citeauthoryear{Crainiceanu, Staicu, and Di}{Crainiceanu
  et~al.}{2009}]{Crainiceanu:09}
Crainiceanu, C., A.~Staicu, and C.~Di (2009).
\newblock {Generalized Multilevel Functional Regression}.
\newblock {\em J. Amer. Statist. Assoc.\/}~{\em 104}, 1550--1561.

\bibitem[\protect\citeauthoryear{Crainiceanu, Staicu, Ray, and
  Punjabi}{Crainiceanu et~al.}{2012}]{Crainiceanu:12}
Crainiceanu, C., A.~Staicu, S.~Ray, and N.~Punjabi (2012).
\newblock {Bootstrap-based inference on the difference in the means of two
  correlated functional processes}.
\newblock {\em Statist. Med.\/}~{\em 31}, 3223--3240.

\bibitem[\protect\citeauthoryear{Craven and Wahba}{Craven and
  Wahba}{1979}]{Wahba:79}
Craven, P. and G.~Wahba (1979).
\newblock {Smoothing noisy data with spline functions}.
\newblock {\em Numer. Math.\/}~{\em 31}, 377--403.

\bibitem[\protect\citeauthoryear{Cummins, Filloon, and Nychka}{Cummins
  et~al.}{2001}]{Cummins:01}
Cummins, D., T.~Filloon, and D.~Nychka (2001).
\newblock {Confidence intervals for nonparametric curve estimates: toward more
  uniform pointwise coverage}.
\newblock {\em J. Amer. Stat. Assoc.\/}~{\em 96}, 233--246.

\bibitem[\protect\citeauthoryear{Dauxois, Pousse, and Romain}{Dauxois
  et~al.}{1982}]{Dauxois:82}
Dauxois, J., A.~Pousse, and Y.~Romain (1982).
\newblock {Simultaneous nonparametric regressions of unbalanced longitudinal
  data}.
\newblock {\em J. Multivariate Anal.\/}~{\em 12}, 136--154.

\bibitem[\protect\citeauthoryear{Di, Crainiceanu, Caffo, and Punjabi}{Di
  et~al.}{2009}]{Di:09}
Di, C., C.~M. Crainiceanu, B.~S. Caffo, and N.~Punjabi (2009).
\newblock {Multilevel functional principal component analysis}.
\newblock {\em Ann. Appl. Statist.\/}~{\em 3}, 458--488.

\bibitem[\protect\citeauthoryear{Eilers and Marx}{Eilers and
  Marx}{1996}]{Eilers:96}
Eilers, P. and B.~Marx (1996).
\newblock {Flexible smoothing with B-splines and penalties (with Discussion)}.
\newblock {\em Statist. Sci.\/}~{\em 11}, 89--121.

\bibitem[\protect\citeauthoryear{Eilers and Marx}{Eilers and
  Marx}{2003}]{Eilers:03}
Eilers, P. and B.~Marx (2003).
\newblock {Multivariate calibration with temperature interaction using
  two-dimensional penalized signal regression}.
\newblock {\em Chemometrics and Intelligent Laboratory Systems\/}~{\em 66},
  159--174.

\bibitem[\protect\citeauthoryear{Goldsmith, Bobb, Crainiceanu, Caffo, and
  Reich}{Goldsmith et~al.}{2011}]{Goldsmith:11}
Goldsmith, J., J.~Bobb, C.~Crainiceanu, B.~Caffo, and D.~Reich (2011).
\newblock {Longitudinal functional principal component}.
\newblock {\em J. Comput. Graph. Statist.\/}~{\em 20}, 830--851.

\bibitem[\protect\citeauthoryear{Greven, Crainiceanu, Caffo, and Reich}{Greven
  et~al.}{2010}]{Greven:10}
Greven, S., C.~Crainiceanu, B.~Caffo, and D.~Reich (2010).
\newblock {Longitudinal functional principal component}.
\newblock {\em Electronic J. Statist.\/}~{\em 4}, 1022--1054.

\bibitem[\protect\citeauthoryear{Karhunen}{Karhunen}{1947}]{Karhunen:47}
Karhunen, K. (1947).
\newblock {Uber lineare methoden in der wahrscheinlichkeitsrechnung}.
\newblock {\em Annales Academie Scientiarum Fennicae\/}~{\em 37}, 1--79.

\bibitem[\protect\citeauthoryear{Kim and Gu}{Kim and Gu}{2004}]{Kim:04}
Kim, Y.~J. and C.~Gu (2004).
\newblock {Smoothing spline Gaussian regression: more scalable computation via
  efficient approximation}.
\newblock {\em J. R. Statist. Soc. B\/}~{\em 66}, 337--356.

\bibitem[\protect\citeauthoryear{Kneip}{Kneip}{1994}]{Kneip:94}
Kneip, A. (1994).
\newblock {Nonparametric estimation of common regressors for similar curve
  data}.
\newblock {\em Ann. Statist.\/}~{\em 22}, 1386--1427.

\bibitem[\protect\citeauthoryear{Marx and Eilers}{Marx and
  Eilers}{2005}]{Marx:05}
Marx, B. and P.~Eilers (2005).
\newblock {Multidimensional Penalized Signal Regression}.
\newblock {\em Technometrics\/}~{\em 47}, 13--22.

\bibitem[\protect\citeauthoryear{Ramsay and Dalzell}{Ramsay and
  Dalzell}{1991}]{Ramsay:91}
Ramsay, J. and C.~J. Dalzell (1991).
\newblock {Some tools for functional data analysis (with Discussion)}.
\newblock {\em J. R. Statist. Soc. B\/}~{\em 53}, 539--572.

\bibitem[\protect\citeauthoryear{Ramsay and Silverman}{Ramsay and
  Silverman}{2005}]{Ramsay:05}
Ramsay, J. and B.~Silverman (2005).
\newblock {\em Functional data analysis}.
\newblock New York: Springer.

\bibitem[\protect\citeauthoryear{Ramsay and Silverman}{Ramsay and
  Silverman}{2002}]{Ramsay:02}
Ramsay, J. and B.~W. Silverman (2002).
\newblock {\em Applied Functional data analysis: Methods and Case Studies}.
\newblock New York: Springer.

\bibitem[\protect\citeauthoryear{Rice and Silverman}{Rice and
  Silverman}{1991}]{Rice:91}
Rice, J. and B.~Silverman (1991).
\newblock {Estimating the mean and covariance structure nonparametrically when
  the data are curves}.
\newblock {\em J. R. Statist. Soc. B\/}~{\em 53}, 233--243.

\bibitem[\protect\citeauthoryear{Ruppert}{Ruppert}{2002}]{Ruppert:02}
Ruppert, D. (2002).
\newblock {Selecting the number of knots for penalized splines}.
\newblock {\em J. Comput. Graph. Statist.\/}~{\em 1}, 735--757.

\bibitem[\protect\citeauthoryear{Ruppert, Wand, and Carroll}{Ruppert
  et~al.}{2003}]{Ruppert:03}
Ruppert, D., M.~Wand, and R.~Carroll (2003).
\newblock {\em Semiparametric Regression}.
\newblock Cambridge: Cambridge University Press.

\bibitem[\protect\citeauthoryear{Seber}{Seber}{2007}]{Seber:07}
Seber, G. (2007).
\newblock {\em A Matrix Handbook for Statisticians}.
\newblock New Jersey: Wiley-Inter\-sci\-ence.

\bibitem[\protect\citeauthoryear{Shinohara, Crainiceanu, Caffo, and
  Reich}{Shinohara et~al.}{2014}]{Shinohara:14}
Shinohara, R., C.~Crainiceanu, B.~Caffo, and D.~Reich (2014).
\newblock {Longitudinal analysis of spatio-temporal processes: A case study of
  Dynamic Contrast-Enhanced Magnetic Resonance Imaging in Multiple Sclerosis}.
\newblock URL:\url{http://biostats.bepress.com/jhubiostat/paper231/}.

\bibitem[\protect\citeauthoryear{Shou, Zipunnikov, Crainiceanu, and
  Greven}{Shou et~al.}{2013}]{Shou:13}
Shou, H., V.~Zipunnikov, C.~Crainiceanu, and S.~Greven (2013).
\newblock {Structured functional principal component analysis}.
\newblock Available at \url{http://arxiv.org/pdf/1304.6783.pdf}.

\bibitem[\protect\citeauthoryear{Staniswalis and Lee}{Staniswalis and
  Lee}{1998}]{Staniswalis:98}
Staniswalis, J. and J.~Lee (1998).
\newblock {Nonparametric regression analysis of longitudinal data}.
\newblock {\em J. Amer. Statist. Assoc.\/}~{\em 93}, 1403--1418.

\bibitem[\protect\citeauthoryear{Swihart, Caffo, Crainiceanu, and
  Punjabi}{Swihart et~al.}{2012}]{Swihart:12}
Swihart, B., B.~Caffo, C.~Crainiceanu, and N.~Punjabi (2012).
\newblock Mixed effect poisson log-linear models for clinical and
  epidemiological sleep hypnogram data.
\newblock {\em Stat. Med.\/}~{\em 31}, 855--870.

\bibitem[\protect\citeauthoryear{Wang, Shen, and Ruppert}{Wang
  et~al.}{2011}]{Wang:11}
Wang, X., J.~Shen, and D.~Ruppert (2011).
\newblock {Some asymptotic results on generalized penalized spline smoothing}.
\newblock {\em Electronic J. Statist.\/}~{\em 4}, 1--17.

\bibitem[\protect\citeauthoryear{Wood}{Wood}{2003}]{Wood:03}
Wood, S. (2003).
\newblock {Thin plate regression splines}.
\newblock {\em J. R. Statist. Soc. B\/}~{\em 65}, 95--114.

\bibitem[\protect\citeauthoryear{Wood}{Wood}{2013}]{Wood:13}
Wood, S. (2013).
\newblock {R package {\it mgcv}: Mixed GAM computation vehicle with
  GCV/AIC/REML, smoothese estimation (version 1.7-24)}.
\newblock URL:\url{http://cran.r-project.org/web/packages/mgcv/index.html}.

\bibitem[\protect\citeauthoryear{Xiao, Li, Apanasovich, and Ruppert}{Xiao
  et~al.}{2012}]{Xiao:11}
Xiao, L., Y.~Li, T.~Apanasovich, and D.~Ruppert (2012).
\newblock {Local asymptotics of \textsl{P}-splines}.
\newblock Available at \url{http://arxiv.org/abs/1201.0708v3}.

\bibitem[\protect\citeauthoryear{Xiao, Li, and Ruppert}{Xiao
  et~al.}{2013}]{Xiao:12}
Xiao, L., Y.~Li, and D.~Ruppert (2013).
\newblock {Fast bivariate \textsl{P}-splines: the sandwich smoother}.
\newblock {\em J. R. Statist. Soc. B\/}~{\em 75}, 577--599.

\bibitem[\protect\citeauthoryear{Yao, M\"uller, Clifford, Dueker, Follett, Lin,
  Buchholz, and Vogel}{Yao et~al.}{2003}]{Yao:03}
Yao, F., H.~M\"uller, A.~Clifford, S.~Dueker, J.~Follett, Y.~Lin, B.~Buchholz,
  and J.~Vogel (2003).
\newblock {Shrinkage estimation for functional principal component scores with
  application to the population kinetics of plasma folate}.
\newblock {\em Biometrics\/}~{\em 20}, 852--873.

\bibitem[\protect\citeauthoryear{Yao, M\"uller, and Wang}{Yao
  et~al.}{2005}]{Yao:05}
Yao, F., H.~M\"uller, and J.~Wang (2005).
\newblock {Functional data analysis for sparse longitudinal data}.
\newblock {\em J. Amer. Statist. Assoc.\/}~{\em 100}, 577--590.

\bibitem[\protect\citeauthoryear{Zhang and Chen}{Zhang and
  Chen}{2007}]{Zhang:07}
Zhang, J. and J.~Chen (2007).
\newblock {Statistical inferences for functional data}.
\newblock {\em Ann. Statist.\/}~{\em 35}, 1052--1079.

\bibitem[\protect\citeauthoryear{Zipunnikov, Caffo, Crainiceanu, Yousem,
  Davatzikos, and Schwartz}{Zipunnikov et~al.}{2011}]{Zipunnikov:11}
Zipunnikov, V., B.~S. Caffo, C.~M. Crainiceanu, D.~Yousem, C.~Davatzikos, and
  B.~Schwartz (2011).
\newblock {Multilevel functional principal component analysis for
  high-dimensional data}.
\newblock {\em J. Comput. Graph. Statist.\/}~{\em 20}, 852--873.

\bibitem[\protect\citeauthoryear{Zipunnikov, Greven, Caffo, and
  Crainiceanu}{Zipunnikov et~al.}{2012}]{Zipunnikov:12}
Zipunnikov, V., S.~Greven, B.~S. Caffo, and C.~Crainiceanu (2012).
\newblock {Longitudinal high-dimensional data analysis}.
\newblock Available at \url{http://biostats.bepress.com/jhubiostat/paper234/}.

\end{thebibliography}


\begin{table*}
\caption{\label{table_mise_eigenfunction} $100\times$MISEs of the
three methods for estimating the eigenfunctions. The incomplete data
has about 13\% observations missing.} \centering \fbox{
\begin{tabular}{ccccccc}
&\multirow{2}{*}{Eigenfunction}&\multirow{2}{*}{No smoothing}&\multirow{2}{*}{SSVD}&\multirow{2}{*}{S-Smooth}&\multirow{2}{*}{FACE}&FACE\\
&&&&&&incomplete data\\\hline
\multirow{3}{*}{Case 1}&1&9.19&7.27&7.01&6.86&6.97\\
&2&16.95&12.12&11.76&11.65&11.96\\
&3&20.27&6.90&6.74&6.74&6.74\\\hline
\multirow{3}{*}{Case 2}&1&10.05&6.41&6.39&6.29&6.34\\
&2&17.38&11.13&10.92&10.37&10.46\\
&3&19.71&6.75&6.51&6.08&6.23\\\hline
\multirow{3}{*}{Case 3}&1&3.14&0.58&0.58&0.58&0.58\\
&2&23.84&4.40&4.37&4.37&4.37\\
&3&55.51&14.07&13.40&13.41&13.14\\\hline
\multirow{3}{*}{Case 4}&1&5.09&1.81&1.80&1.80&1.87\\
&2&20.14&8.23&8.20&8.20&8.67\\
&3&42.04&19.39&19.39&19.40&20.70\\\hline
\multirow{3}{*}{Case 5}&1&70.34&64.71&64.71&64.71&65.79\\
&2&96.39&90.57&90.31&90.38&90.84\\
&3&93.09&84.15&83.88&83.99&84.66\\
\end{tabular}}
\end{table*}

\begin{table*}
\caption{\label{table_mise_covariance}$100\times$MISEs of the three
methods for estimating the covariance function.  The incomplete data
has about 13\% observations missing. } \centering \fbox{
\begin{tabular}{ccccc}
&\multirow{2}{*}{SSVD}&\multirow{2}{*}{S-Smooth}&\multirow{2}{*}{FACE}&FACE\\
&&&&incomplete data\\
\hline
Case 1&9.34&8.96&8.94&8.93\\
Case 2&8.96&8.64&8.62&8.69\\
Case 3 &1.22&0.76&0.76&0.76\\
Case 4 &0.11&0.07&0.07&0.08\\
Case 5&2.69&1.98&1.98&2.18
\end{tabular}}
\end{table*}

\begin{table*}
\caption{\label{table_mise_eigenvalue} $100\times$ average
$(\hat\lambda_k/\lambda_k-1)^2$ of the three methods for estimating
the eigenvalues. The incomplete data has about 13\% observations
missing.} \centering \fbox{
\begin{tabular}{cccccc}
&\multirow{2}{*}{Eigenvalue}&\multirow{2}{*}{SSVD}&\multirow{2}{*}{S-Smooth}&\multirow{2}{*}{FACE}&FACE\\
&&&&&incomplete data\\\hline
\multirow{3}{*}{Case 1}&1&4.37&3.99&3.99&4.31\\
&2&3.43&3.68&3.76&3.96\\
&3&3.97&4.95&5.03&4.99\\\hline
\multirow{3}{*}{Case 2}&1&4.40&4.05&4.05&4.10\\
&2&3.58&3.78&3.81&3.83\\
&3&3.38&4.02&4.38&4.22\\\hline
\multirow{3}{*}{Case 3}&1&3.80&3.55&3.55&3.55\\
&2&9.79&3.38&3.38&3.42\\
&3&48.27&4.03&4.03&3.96\\\hline
\multirow{3}{*}{Case 4}&1&4.22&3.81&3.81&3.84\\
&2&5.65&3.69&3.69&3.64\\
&3&14.77&3.53&3.53&3.43\\\hline
\multirow{3}{*}{Case 5}&1&12.45&6.45&6.45&7.05\\
&2&4.35&2.09&2.09&2.03\\
&3&3.05&1.64&1.64&1.55\\
\end{tabular}}
\end{table*}

\begin{table*}
\caption{\label{time}Computation time (in seconds) of the SSVD,
S-Smooth and FACE methods averaged over 100 data sets on 2.4GHz Mac
computers  with 8 gigabytes of random access memory.  The
computation time of the sandwich smoother is also provided except
for $J=10,000$ and is averaged over 10 datasets only. } \centering
\fbox{
\begin{tabular}{cccccccc}
\multirow{2}{*}{$J$}&\multirow{2}{*}{$I$}&\multirow{2}{*}{SSVD}&\multirow{2}{*}{S-Smooth}&FACE&FACE&Sandwich&Sandwich\\
&&&&$100$ knots&$500$ knots &$100$ knots&$500$ knots\\\hline
\multirow{2}{*}{$3,000$}&$50$&0.25&1.28&0.34&1.76&47.41&210.41\\
&$500$&3.81&13.88&0.89&2.61&50.91&364.39\\\hline
\multirow{2}{*}{$5,000$}&$50$&0.43&2.14&0.50&2.09&251.48&1362.67\\
&$500$&6.08&34.63&1.26&3.19&302.34&1743.86\\\hline
\multirow{2}{*}{$10,000$}&$50$&0.86&4.29&0.82&2.92&-&-\\
&$500$&12.78&98.41&2.34&4.68&-&-\\
\end{tabular}}
\end{table*}
\clearpage

\renewcommand{\arraystretch}{1.5}
\begin{table*}
\caption{\label{eigenvalues} Estimated eigenvalues of $K_X$ and
$K_U$. All eigenvalues  are multiplied by $J$ to refer to the
variation in the data explained by the eigenfunctions. The row `all'
refers to the sum of all positive eigenvalues.} \centering \fbox{
\begin{tabular}{ccccc}
&Eigenfunction&SSVD&FACE&Thin Plate Splines\\\hline
\multirow{4}{*}{$K_X$}&1&4.31&3.92&1.91\\
 &2&2.64&2.66&0.50\\
 &3&1.88&1.35&0.31\\
 &all&48.14&14.40&2.81\\\hline
\multirow{4}{*}{$K_U$}&1&8.84&6.33&6.75\\
 &2&5.69&3.18&2.55\\
 &3&5.03&2.86&2.04\\
 &all&107.95&22.75&12.95\\
\end{tabular}}
\end{table*}

\clearpage

\end{document}